\documentclass[aps,prl,twocolumn,groupedaddress]{revtex4}

\usepackage{graphicx}
\usepackage{dcolumn}
\usepackage{bm}
\usepackage{subfigure}

\begin{document}


\title{Comment on "Conformal invariance and stochastic Loewner
evolution processes in two-dimensional Ising spin glasses"}


\author{Ronald Fisch}
\affiliation{382 Willowbrook Dr.\\
North Brunswick, NJ 08902}


\date{\today}

\begin{abstract}
By combining the scaling relation of Amoruso {\it et al.}, PRL {\bf
97}, 267202 (2006) with standard droplet model assumptions, a value
$\theta = (\sqrt{6} - 3) / 2$ is obtained.  This conjecture is
reasonably consistent with the best existing numerical calculations,
and may be exact.

\end{abstract}


\maketitle

Recently, Amoruso {\it et al.}\cite{AHHM06} have given an argument
for a scaling relation
\begin{equation}
  d_f = 1 + {3 \over {4 ( 3 + \theta )}}   \,
\end{equation}
in the two-dimensional Ising spin glass\cite{EA75} with continuous
({\it e.g.} Gaussian) bond distributions.  In Eqn.~(1), $d_f$ is the
fractal dimension of domain walls, and $\theta$ is the finite-size
scaling exponent for the energy of the domain walls.  In this
Comment we wish to point out that by combining this relation with
standard results from a droplet scaling analysis for this
model,\cite{LM07} an exact value for the exponent ${\theta}$ is
obtained.

From the droplet model,\cite{LM07} we obtain a simple relation
\begin{equation}
  d_S = d/2 - \theta   \, ,
\end{equation}
where $d_S$ is the finite-size scaling exponent for droplets, and
$d$ is the number of dimensions of space ({\it i.e.} $d = 2$).
Within the spirit of the droplet model, all lengths should scale
with the same exponent.  Therefore, if this assumption is valid, we
are justified in setting $d_S = d_f$.  Combining Eqn.~1 and Eqn.~2,
we then have a quadratic equation for $\theta$.  The relevant
solution is the larger one,
\begin{equation}
  \theta = (\sqrt{6} - 3) / 2 = -0.275255... \, .
\end{equation}
Eqn.~3 is reasonably consistent with recent estimates of $\theta$
for this model.\cite{Har07}  Therefore, we conjecture that it is
exact.

According to the droplet model the thermal exponent $\nu$ is
supposed to be given by $\nu = -1 / \theta$, so we also find
\begin{equation}
  \nu = 2 /(3 - \sqrt{6}) = 3.63299... \, .
\end{equation}

\begin{acknowledgments}
The author thanks the Physics Department of Princeton University for
providing use of facilities.

\end{acknowledgments}



\end{document}